\documentclass[lettersize,journal]{IEEEtran}
\usepackage{amsmath,amsfonts}
\usepackage{algorithmic}
\usepackage{algorithm}
\usepackage{array}
\usepackage[caption=false,font=normalsize,labelfont=sf,textfont=sf]{subfig}
\usepackage{textcomp}
\usepackage{stfloats}
\usepackage{url}
\usepackage{verbatim}
\usepackage{graphicx}
\usepackage{cite}
\usepackage{xcolor}
\usepackage{balance}
\begin{document}

\title{Simultaneous Classical and Quantum Communications: Recent Progress and Three Challenges}

\author{Phuc V. Trinh,~\IEEEmembership{Senior Member,~IEEE,} Shinya Sugiura,~\IEEEmembership{Senior Member,~IEEE,} Carlo Ottaviani, Chao~Xu,~\IEEEmembership{Senior Member,~IEEE,} and Lajos~Hanzo,~\IEEEmembership{Life Fellow,~IEEE}
        % <-this % stops a space
\thanks{Preprint (Accepted Version). DOI: 10.1109/MNET.2025.3643793.
Copyright $\copyright$ 2025 IEEE. Personal use of this material is permitted. However, permission to use this material for any other purposes must be obtained from the IEEE by sending a request to pubs-permissions@ieee.org.}
\thanks{Phuc V. Trinh and Shinya Sugiura are with the Institute of Industrial Science, The University of Tokyo, Tokyo 153-8505, Japan; Carlo Ottaviani is with the Department of Computer Science and Centre for Quantum Technologies, University of York, YO10 5GH York, U.K.; Chao Xu and Lajos Hanzo (corresponding author) are with the School of Electronics and Computer Science, University of Southampton, SO17 1BJ Southampton, U.K.}
}

\markboth{Preprint (Accepted Version) for publication in IEEE Network (DOI: 10.1109/MNET.2025.3643793)}%
{Trinh \MakeLowercase{\textit{et al.}}: Simultaneous Classical and Quantum Communications: Recent Progress and Three Challenges}


\maketitle

\begin{abstract}
A critical aspect of next-generation wireless networks is the integration of quantum communications to guard against quantum computing threats to classical networks. Despite successful experimental demonstrations, integrating quantum communications into the classical infrastructure faces substantial challenges, including high costs, compatibility issues, and extra hardware deployment to accommodate both classical and quantum communication equipment. To mitigate these challenges, we explore novel protocols that enable simultaneous classical and quantum communications, relying on a single set of transceivers to jointly modulate and decode classical and quantum information onto the same signal. Additionally, we emphasize extending quantum communication capabilities beyond traditional optical bands into the terahertz, even possibly to millimeter-wave and microwave frequencies, thereby broadening the potential horizon of quantum-secure applications. Finally, we identify open problems that must be addressed to facilitate practical implementation.
\end{abstract}

%\begin{IEEEkeywords}
%
%\end{IEEEkeywords}
%%---------I. INTRODUCTION-----------
\section{Introduction}
\IEEEPARstart{F}{uture} wireless networks are expected to rely on a space-air-ground integrated network (SAGIN) architecture, seamlessly connecting the terrestrial infrastructure with diverse aerial and space-based platforms, including unmanned aerial vehicles (UAVs), high-altitude platforms (HAPs), and Low-Earth Orbit (LEO) satellite constellations \cite{Khan2025}. These advanced networks intelligently leverage multiple frequency bands across the electromagnetic spectrum, spanning the microwave, millimeter wave (mmWave), terahertz (THz), and free-space optical (FSO) frequencies, to ensure robust and adaptive connectivity among various network layers. This multi-layer and multi-band integration is expected to significantly enhance network resilience, support high-capacity data transmission, and enable reliable ubiquitous global Internet connectivity across diverse applications. 

Building upon the SAGIN concept, a recent study has outlined a vision of the quantum Internet (Qinternet) relying on non-terrestrial networks utilizing FSO systems \cite{Trinh2024}. The envisioned \textit{Qinternet in the sky} focuses on manipulating quantum states within optical frequencies for distributing cryptographic keys via quantum key distribution (QKD) protocols and establishing entanglement links among distant SAGIN nodes. This facilitates quantum information transfer between remote end nodes without requiring continuous direct physical connections, as entanglement can be distributed and stored across the network using quantum memories and previously shared entangled photons. Nevertheless, these FSO links face inherent line-of-sight constraints and practical limitations associated with communication terminals mounted beneath UAVs. To overcome these challenges, optical reconfigurable intelligent surfaces (ORISs) relying on passive reconfigurable metasurfaces have been proposed for deployment on building rooftops to assist FSO-based quantum SAGIN communications \cite{Trinh2025}. ORISs enable three-dimensional beam steering toward the targeted non-terrestrial platforms, with beamwidth control optimized to reduce power loss and improve reception efficiency. Based on this foundation, a cooperative coexistence of classical and quantum SAGINs is anticipated.

Currently, SAGIN architectures conceived for classical and quantum communications typically require separate sets of hardware transceivers to support the two signal types \cite{Khan2025,Trinh2024,Trinh2025,Li2025}. This results in either two distinct SAGIN designs or a single SAGIN infrastructure with separate classical and quantum components. However, this duplication poses significant challenges for non-terrestrial platforms, particularly for UAVs and small satellites, where strict constraints on size, weight, power, and cost (SWaP-C) must be observed. Integrating separate transceivers for classical and quantum functions would substantially increase deployment costs and power consumption, while also exceeding the physical and energy limitations of compact airborne or spaceborne platforms. Given that these platforms are generally battery-powered and must support extended operation with minimal resource overhead, adherence to SWaP-C requirements is essential to ensure scalability, long mission duration, and widespread deployment on a global scale.

To address the stringent SWaP-C constraints of non-terrestrial platforms, simultaneous classical and quantum communication (SCQC) schemes present a compelling solution \cite{Pan2025,Qi2016,Kumar2019}. Unlike traditional architectures that require separate transceivers for classical and quantum signals, SCQC enables the transmission of quantum and classical information over a single channel using a common set of transceivers. This is achieved by intelligently encoding classical bits and quantum states onto the same signal, which can then be decoded by a unified receiver. Such integration significantly reduces hardware redundancy and aligns more effectively with the size, weight, and power limitations of small UAVs and satellites. Moreover, simulation studies show that SCQC can maintain low bit error rates for classical communications, while ensuring secure key distribution over practical distances. Beyond optical frequencies \cite{Pan2025,Qi2016,Kumar2019}, a recent breakthrough has demonstrated the feasibility of implementing QKD in the microwave band \cite{Fesquet2024}, while theoretical studies have extended this exploration to higher-frequency regimes such as mmWave and THz \cite{Ottaviani2020,Zhang2023}, thereby adding the potential deployment scenarios for SCQC systems. Integrating SCQC into multi-band SAGIN architectures thus offers a promising path toward cost-effective, scalable, and quantum-secure wireless futures. Nevertheless, bringing SCQC closer to practical deployment certainly requires a clear identification of key open challenges and dedicated efforts to develop viable solutions.
%%---------II. SCQC for Multi-Band SAGINs-----------
\section{SCQC for Multi-Band SAGINs}
%%--Table 1--
\begin{table*}[t]
\centering
\captionsetup{font=footnotesize}
\caption{Comparison of simultaneous and coexistent classical and quantum communication schemes.}
\scalebox{1}{%
\begin{tabular}{l|l|l|}
\cline{2-3}
                      & \textbf{Simultaneous schemes} \cite{Pan2025,Qi2016,Kumar2019} & \textbf{Coexistent schemes} \cite{Xu2023,Shao2025,Schreier2023} \\ \hline\hline
\multicolumn{1}{|l|}{\textbf{Channel separation}} & \begin{tabular}[t]{@{}l@{}}\textbf{Fully overlapped}\\(same pulse and time slot)\end{tabular} & \begin{tabular}[t]{@{}l@{}}\textbf{Separate} \\(in wavelength, time, and frequency)\end{tabular}   \\ \hline
\multicolumn{1}{|l|}{\textbf{Hardware}} & \begin{tabular}[t]{@{}l@{}}\textbf{One set} \\ of transmitter/receiver\end{tabular} & \begin{tabular}[t]{@{}l@{}}\textbf{Two sets}\\of transmitters/receivers\end{tabular}   \\ \hline
\multicolumn{1}{|l|}{\textbf{Implementation cost}} & \begin{tabular}[t]{@{}l@{}}\textbf{Low}\\(due to shared hardware)\end{tabular} & \begin{tabular}[t]{@{}l@{}}\textbf{High}\\(due to multiple devices, filters)\end{tabular} \\ \hline
\multicolumn{1}{|l|}{\textbf{Spectral/time efficiency}} & \begin{tabular}[t]{@{}l@{}}\textbf{High}\\(no dedicated guard resources)\end{tabular} & \begin{tabular}[t]{@{}l@{}}\textbf{Low}\\(guard bands, interleaving gaps)\end{tabular}  \\ \hline
\multicolumn{1}{|l|}{\textbf{Impairments}} & \begin{tabular}[t]{@{}l@{}}\textbf{Intrinsic noise} \\(from combined classical-quantum modulation)\end{tabular} & \begin{tabular}[t]{@{}l@{}}\textbf{Crosstalk} \\(classical leakage into the quantum channel)\end{tabular} \\ \hline
\multicolumn{1}{|l|}{\textbf{Classical data rate}} & \begin{tabular}[t]{@{}l@{}}\textbf{Low}\\(limited by detection bandwidth of the quantum receiver) \end{tabular} & \begin{tabular}[t]{@{}l@{}}\textbf{High}\\(comparable to standalone classical rates)\end{tabular}   \\ \hline
\multicolumn{1}{|l|}{\textbf{QKD key rate}} & \begin{tabular}[t]{@{}l@{}}\textbf{Low}\\(limited by detection bandwidth of the quantum receiver) \end{tabular} & \begin{tabular}[t]{@{}l@{}}\textbf{High}\\(comparable to standalone QKD rates)\end{tabular}   \\ \hline
\multicolumn{1}{|l|}{\textbf{Security proof complexity}}  & \begin{tabular}[t]{@{}l@{}}\textbf{Complex}\\(joint noise/security analysis required)\end{tabular} & \begin{tabular}[t]{@{}l@{}}\textbf{Simple}\\(standard quantum security proof)\end{tabular}  \\ \hline
\multicolumn{1}{|l|}{\textbf{Technological maturity}} & \begin{tabular}[t]{@{}l@{}}\textbf{Moderate}\\(experimental demonstration) \end{tabular} & \begin{tabular}[t]{@{}l@{}}\textbf{Advanced}\\(commercial systems available)\end{tabular}   \\ \hline
\end{tabular}%
}
\label{Table_1} 
\end{table*}
%%--------
The objectives of classical and quantum communications in wireless networks are fundamentally complementary. Classical communications enables high-speed data transmission and supports the vast plethora of modern digital services, while quantum communications is generally designed to ensure information-theoretic security based on the principles of quantum mechanics. As the capabilities of quantum computing continue to advance, they pose a serious threat to widely used public-key cryptographic methods, such as the Rivest-Shamir-Adleman algorithm. Despite the promise of quantum communications in addressing this vulnerability, a major practical limitation is the requirement for additional dedicated hardware transceivers and separate communication channels to support quantum-secured protocols. This requirement substantially increases both the capital expenditure and energy consumption, which discourages telecommunication operators and service providers from adopting quantum technologies, particularly in large-scale resource-constrained SAGINs.

To address these concerns, researchers have proposed two main schemes for facilitating the integration of quantum communications into existing classical infrastructures: coexistent classical and quantum communications (CCQC) and SCQC. More specifically, CCQC schemes achieve physical-layer coexistence by allocating classical and quantum signals to distinct time slots \cite{Xu2023}, frequency bands \cite{Shao2025}, or wavelengths \cite{Schreier2023}. Although this approach allows both signal types to share the same physical transmission medium, it still requires separate transceiver hardware for the classical and quantum functions. As a result, CCQC tends to incur higher implementation costs and reduced spectral or time efficiency due to the need for guard bands or interleaving gaps to avoid crosstalk. By contrast, SCQC schemes represent a more tightly integrated approach by jointly modulating classical and quantum information onto the same signal and transmitting them over the same physical channel using a single set of transceivers \cite{Pan2025,Qi2016,Kumar2019}. This integrated infrastructure potentially improves hardware reuse and energy efficiency, making it attractive for deployment in resource-constrained platforms such as satellites and UAVs. However, SCQC faces increased system complexity, primarily due to the noise introduced by signal overlap, which erodes the signal-to-noise ratio (SNR) and may impact both classical and quantum performance. Additionally, SCQC requires more rigorous security analysis, as encoding and decoding classical and quantum information on the same carrier may affect the security guarantees of the quantum protocol.

Table \ref{Table_1} summarizes the differences between SCQC and CCQC, providing a structured overview of important indicators such as hardware requirements, spectral efficiency, interference effects, data and key rates, plus technological maturity. While each scheme presents its own advantages and limitations, SCQC offers a forward-looking strategy for achieving cost-effective and resource-efficient integration with modest modifications to existing infrastructure, making it attractive for synergistic classical and quantum SAGINs.

%%---FIG 1-----
\begin{figure*}[t]
\centering
\includegraphics[scale=0.52]{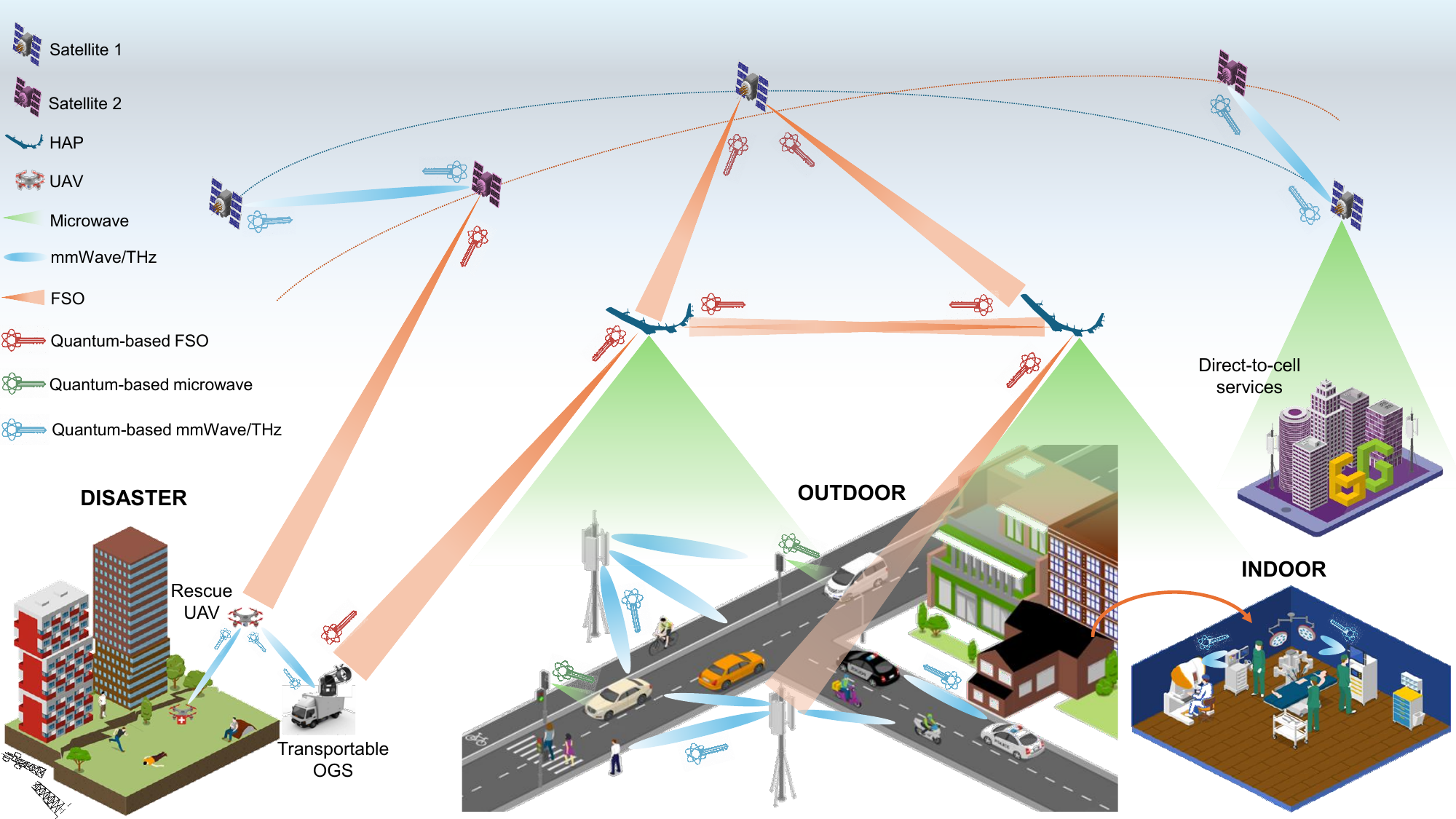}
\caption{Illustration of simultaneous classical and quantum communications for wireless futures.}
\label{fig_1}
\end{figure*}
%%--------
Figure \ref{fig_1} illustrates a vision of multi-band SAGINs empowered by SCQC, highlighting a variety of application scenarios across diverse frequency bands and communication ranges. In this architecture, FSO-based SCQC links \cite{Pan2025,Qi2016,Kumar2019} serve as the optical wireless backbone for long-distance, high-security communications among satellites, HAPs, and transportable optical ground stations (OGS), capitalizing on the limited beam divergence and low interference susceptibility of optical transmissions. In addition, SCQC can be extended beyond the optical spectrum, as quantum cryptography has been explored over microwave \cite{Fesquet2024}, mmWave and THz bands \cite{Ottaviani2020, Zhang2023}, which are well suited for short-range applications leveraging the existing wireless infrastructure and supporting compact hardware integration. These include indoor environments such as secure, low-latency communication for robotic surgery, and outdoor scenarios involving short-range links between small-cell base stations and mobile devices, vehicles, or among vehicles in intelligent transportation systems. However, despite these promising use cases, QKD at longer wavelengths faces significant physical challenges. While QKD in the optical band benefits from low thermal noise, mature photonic technology, and minimal atmospheric attenuation, its counterparts in the microwave regime suffer from strong thermal background noise, severe free-space path loss, and immature quantum hardware, which currently restrict secure transmission to very short distances. On the other hand, the mmWave and THz bands bridge the gap between optical and microwave frequencies, offering reduced thermal noise and narrower beam divergence, while maintaining a moderate level of alignment tolerance. This makes them particularly attractive as alternatives to FSO for enabling limited-range inter-satellite SCQC in LEO. 

In emergency response settings, such as post-earthquake search and rescue operations, SCQC enables reliable, secure communications among autonomous drones and field teams, especially when the terrestrial communication infrastructure is damaged. Integrating SCQC across multiple frequency bands enables SAGINs to seamlessly support both classical and quantum communications within a unified, cost-effective, and resource-efficient framework. The direct-to-cell scenario of Fig. \ref{fig_1} represents another significant advance in future SAGINs, enabling satellites to communicate directly with mobile devices over microwave and sub-$6$ Gigahertz (GHz) cellular bands without relying on terrestrial relay stations. This capability not only ensures ubiquitous connectivity in remote or disaster-stricken regions, but also enhances network coverage, reliability, and mobility support in dense urban environments. Although implementing QKD directly over these links appears infeasible due to severe free-space path loss and quantum decoherence, the direct-to-cell link can still play a crucial role in supporting the classical communications required for QKD postprocessing, including reconciliation, authentication, and key management. This makes it an essential element in future quantum-secure SAGINs.
%%---------III. SCQC over Optical Band-----------
\section{SCQC over Optical Bands}
Recent advances in SCQC have primarily focused on optical frequencies, targeting the simultaneous transmission of classical data and QKD over shared infrastructures such as optical fiber networks \cite{Pan2025,Qi2016,Kumar2019}. QKD protocols allow two parties to share a secret key with information-theoretic security guaranteed by the laws of quantum mechanics. As classical cryptographic methods that rely on mathematical complexity face growing threats from quantum computing, the integration of QKD into communication networks is becoming increasingly critical. Two notable approaches exemplify this effort. 

A promising technique is based on a one-way quasi–quantum secure direct communication (QSDC) protocol that enables the simultaneous transmission of encrypted classical data and quantum key exchange \cite{Pan2025}. This method transmits quantum states that carry classical messages encrypted with a preshared one-time-pad key. Although the protocol depends on classical key material for encryption, it includes a built-in mechanism to re-generate a secure key sequence from a copy of the raw data obtained after quantum measurements, thereby partially replenishing the key sink. Specifically, it employs repeatable encoding, where the same quantum state is used not only to deliver the encrypted message but also, under low-noise and low-loss conditions, to extract new secret key bits from the measurement results. These bits are added to the key sink, helping to sustain secure communications over time. The system supports discrete-variable (DV) QKD by incorporating single-photon-level encoding along with decoy-state techniques to ensure the security of the quantum channel. This approach has been experimentally demonstrated over standard telecommunication fiber across a distance exceeding $100$ kilometers (km), achieving secure transmission rates suitable for applications such as text, image files, and voice communication \cite{Pan2025}. This enables existing DV QKD systems to support classical data transmission as an additional feature, using the same optical channel and transceiver hardware for both secure key exchange and classical communications.

A second approach supports simultaneous classical communications and continuous-variable (CV) QKD by encoding both classical bits and quantum key material onto the same coherent states. This is possible because the classical coherent communication signal exhibits quantum uncertainty at intermediate signal strengths, allowing both classical and quantum information to be conveyed in a single mode. Classical data is embedded using phase-shift keying by displacing the phase of the coherent state, while quantum information is encoded as Gaussian-modulated quadrature variables \cite{Qi2016}. This combined encoding produces displaced Gaussian-modulated coherent states (GMCS), where the classical bit determines the direction of displacement in phase space. At the receiver, a coherent detection system such as homodyne or heterodyne detection mixes the incoming signal with a strong local oscillator (LO). This setup enables accurate quadrature measurements using photodiodes and inherently filters broadband noise photons generated in the communication channel. The classical bit is recovered by taking the sign of the measured quadrature value after the phase correction, while the same measurement result is rescaled and corrected to extract the quantum Gaussian variable used for key generation \cite{Qi2016}. This scheme builds on the GMCS protocol, leveraging its compatibility with classical coherent systems to enable joint transmission and decoding of classical and quantum signals over a shared optical channel using common transceivers. Experimental demonstrations have shown successful operation over $25$ km of standard telecommunication fiber \cite{Kumar2019}, confirming its practicality for secure and efficient integration of quantum functionality into existing classical infrastructure.

SCQC schemes originally developed for optical fiber links in \cite{Pan2025,Qi2016} can also be directly applied to FSO systems. As a representative use case for SAGINs, we consider a LEO satellite-to-ground FSO downlink that accounts for diffraction, refraction, atmospheric extinction, pointing errors, and turbulence. Zenith-crossing satellites at various altitudes are analyzed, focusing on the worst-case condition with a zenith angle of 80 degrees, corresponding to moderate-to-strong turbulence and representing the maximum acceptable angle to maintain an unobstructed line of sight. This condition provides a lower bound on the overall system performance. The composable key rates of both DV and CV QKD schemes in \cite{Pan2025,Qi2016} are evaluated following the frameworks in \cite{Vasylyev2019,Ghalaii2022}. The composable key rate provides a rigorous finite-size security measure that quantifies how efficiently a QKD protocol can generate a key that remains secure against general coherent attacks. Within this framework, privacy amplification distills a shorter but fully secret key, ensuring that the final key remains secure even when integrated into larger cryptographic applications such as encryption or authentication.

%%---Fig. 2------
\begin{figure}[t]
\centering
\includegraphics[scale=0.5]{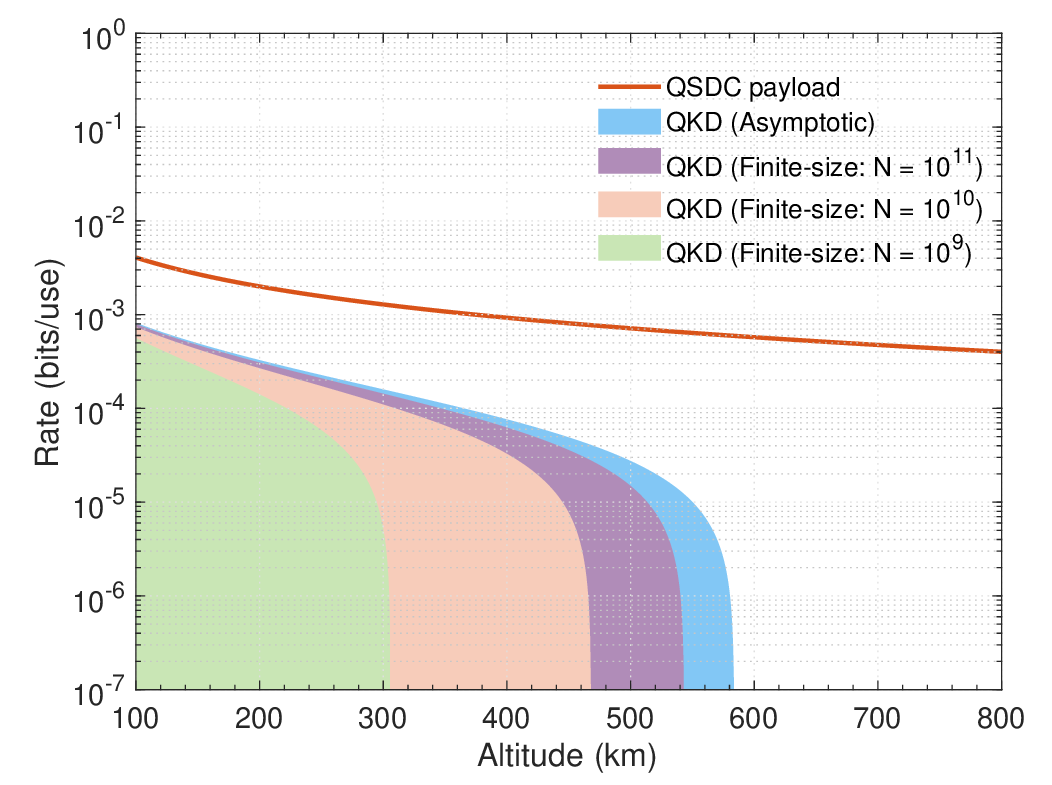}
\caption{Rate performance of the SCQC scheme employing DV QKD as described in \cite{Pan2025} for a satellite-to-ground FSO downlink. The FSO channel characteristics are adopted from \cite{Ghalaii2022}, with key parameters including a zenith angle of 80 degrees, an initial spot size of $20$ centimeters (cm), a receiver aperture radius of $70$ cm, and an optical wavelength of $800$ nanometers (nm). The composable key rate evaluation follows the parameter settings in \cite{Pan2025,Vasylyev2019}, with main parameters including a mean signal intensity of $0.6$, a mean weak-decoy intensity of $0.2$, a receiver efficiency of $20\%$, a background error rate of $0.5$, a failure probability of $10^{-10}$, a background yield due to stray light of $2\times10^{-4}$, a background yield due to dark counts of $2.4\times10^{-6}$, and an error-correction efficiency of $1.05$.}
\label{fig_2}
\end{figure}
%%-------------
%%---Fig. 3------
\begin{figure}[t]
\centering
\includegraphics[scale=0.5]{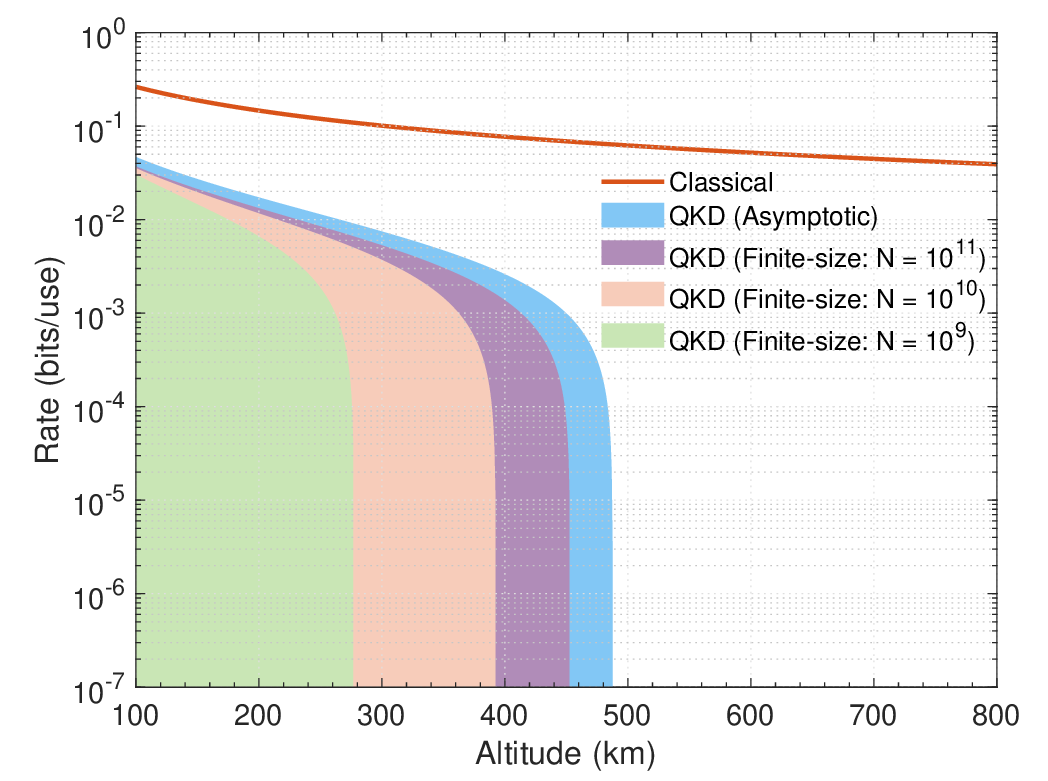}
\caption{Rate performance of the SCQC scheme employing CV QKD as described in \cite{Qi2016} for a satellite-to-ground FSO downlink. The FSO channel characteristics are adopted from \cite{Ghalaii2022}, with key parameters consistent with those in Fig. \ref{fig_2}. The composable key rate evaluation follows the parameter settings in \cite{Qi2016,Ghalaii2022}, with main parameters including a coherent-state variance of $5$, a detector noise of $0.1$, a shot-noise variance of $0.25$, a receiver efficiency of $50\%$, a local LO-induced loss of $0.63$, background thermal photons per mode of $9.31\times10^{-10}$, target classical bit-error rate of $10^{-6}$, a success probability of error correction of $0.9$, a correctness bounding of $10^{-10}$, and a reconciliation efficiency of $0.98$.}
\label{fig_3}
\end{figure}
%%-------------
%
Figures \ref{fig_2} and \ref{fig_3} show the performance evaluation of the SCQC schemes in \cite{Pan2025} and \cite{Qi2016}, respectively, for satellite-to-ground FSO downlinks. In Fig. \ref{fig_2}, the composable key rate and corresponding secure satellite altitudes of the two-decoy-state DV QKD protocol are analyzed for various finite-size data blocks denoted by $N$. The asymptotic case with an infinite block size is also shown as a benchmark. Under the given parameter settings, the results indicate that a LEO satellite can achieve a positive key rate up to an altitude of $583$ km in the asymptotic limit, which then decreases to $542$ km, $467$ km, and $305$ km for $N=10^{11}$, $N=10^{10}$, and $N=10^{9}$, respectively. The corresponding QSDC payload rates are $5.9\times10^{-4}$, $6.4\times10^{-4}$, $7.7\times10^{-4}$, and $1.2\times10^{-3}$ bits/use, evaluated at the respective secure operating altitudes. Here, the term “QSDC payload rate” refers to the net rate of secure message bits transmitted via quantum states, accounting for resources used in eavesdropping detection, synchronization, and error correction, rather than the raw throughput of a conventional classical channel. Although the maximum secure altitude decreases with smaller finite block sizes, the higher payload rates correspond to shorter operating distances where the channel quality and SNR are more favorable. These results confirm that the SCQC scheme in \cite{Pan2025} remains feasible for satellite-to-ground downlinks under the examined conditions.

Figure \ref{fig_3} presents the corresponding results for the GMCS CV QKD scheme in \cite{Qi2016}, evaluated under the same satellite-to-ground FSO channel settings as in Fig. \ref{fig_2}. A LEO satellite can achieve a positive composable key rate up to an altitude of $487$ km in the asymptotic limit, which decreases to $452$ km, $392$ km, and $276$ km when $N=10^{11}$, $N=10^{10}$, and $N=10^{9}$, respectively. The corresponding classical information rates are $6.3\times10^{-2}$, $6.8\times10^{-2}$, $7.8\times10^{-2}$, and $1.08\times10^{-1}$ bits/use, evaluated at the respective secure operating altitudes. The classical rate represents the Shannon mutual information between the legitimate transmitter and receiver, incorporating the reconciliation efficiency. The composable key analysis follows the thermal-loss model of \cite{Ghalaii2022}. Since \cite{Qi2016} considers fiber-based channels and expresses the excess noise from classical phase encoding in shot-noise units, we convert this contribution into an equivalent number of thermal photons following \cite{Ghalaii2022}. Consistently, the trend of decreasing secure altitude and increasing classical rate at shorter distances agrees with Fig. \ref{fig_2}. Finally, these results indicate that the SCQC scheme in \cite{Qi2016} remains feasible for satellite-to-ground optical links under realistic conditions.

The performance differences between the schemes in Figs. \ref{fig_2} and \ref{fig_3} reflect the inherent trade-offs between DV and CV QKD protocols and are presented as a qualitative rather than a quantitative comparison. The DV scheme, based on single-photon detection and decoy-state techniques, achieves longer secure distances by tolerating high channel loss and effectively bounding excess noise through photon-number estimation, enabling secure operation at very low photon flux levels. By contrast, the CV scheme employs homodyne detection with coherent-state modulation. When the SNR is sufficiently high, its continuous encoding allows more mutual information to be extracted from each channel use, resulting in higher achievable key rates. Moreover, CV QKD systems employ coherent light sources, phase modulators, and balanced photodiodes that are standard in classical optical communications, providing broad bandwidth and natural compatibility with existing telecommunication infrastructure, thereby facilitating large-scale integration. However, CV QKD is more sensitive to channel loss and excess noise, particularly in the CV-based SCQC scheme of \cite{Qi2016}, where classical phase modulation implemented by displacing the coherent states introduces additional excess noise that degrades overall system performance.
%%%
%%---------IV. SCQC over mmWave/THz Bands-----------
\section{SCQC over Microwave/mmWave/THz Bands}
Extending SCQC beyond optical frequencies into the microwave, mmWave, and THz bands requires a clear understanding of atmospheric propagation effects, particularly of the attenuation caused by molecular absorption. These losses vary with both frequency and altitude, and can significantly constrain system performance. According to ITU-R P.676, attenuation across the 1 to 1000 GHz range is highly frequency-dependent, and for SAGINs, it can be evaluated along the slant path at an arbitrary elevation angle, taking into account altitude-dependent variations in temperature, pressure, and water-vapor density. 

%%---Fig. 4------
\begin{figure}[t]
\centering
\includegraphics[scale=0.55]{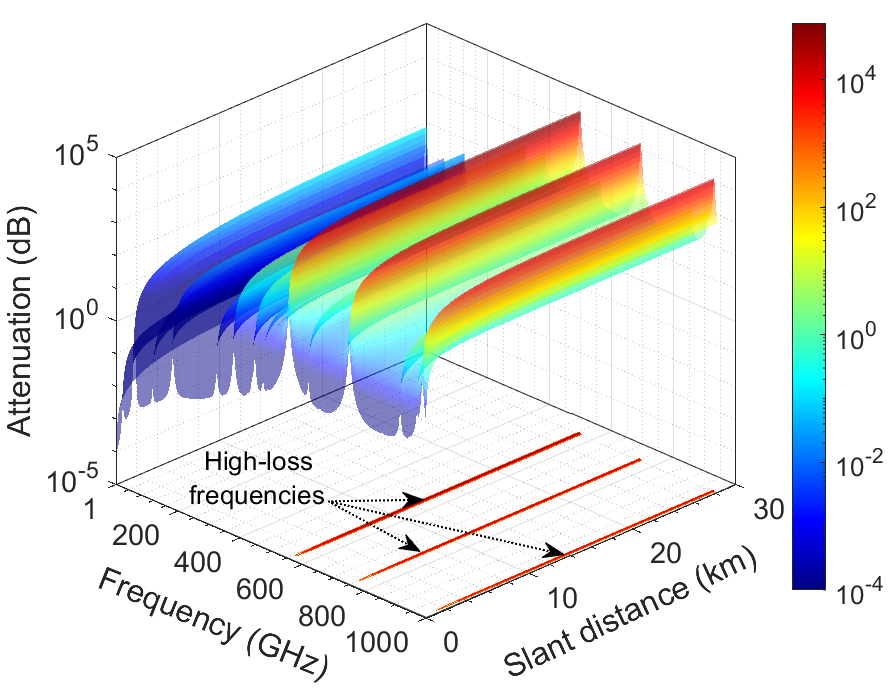}
\caption{Altitude-dependent atmospheric gas attenuation as a function of frequency and slant distance for an elevation angle of 45 degrees.}
\label{fig_4}
\end{figure}
%%-------------
Figure \ref{fig_4} illustrates how attenuation varies with frequency and slant distance for an elevation angle of 45 degrees. In the microwave band (1 to 30 GHz), losses remain relatively low even over slant distances approaching 28 km. In the mmWave region (30 to 300 GHz), attenuation becomes more prominent at specific frequencies, requiring careful spectral selection. In the THz region (above 300 GHz), attenuation increases rapidly with frequency, limiting feasible slant communication distances to below 1 km. For SCQC deployment over mmWave and THz bands, it is important to operate within atmospheric transmission windows where attenuation remains low. This is especially critical in the THz range, where high-loss regions near 557 GHz, 752 GHz, and 988 GHz should be avoided, as shown in Fig. \ref{fig_4}. The design of SCQC systems in these frequency bands, therefore, requires a careful balance between spectral availability and propagation loss, highlighting the need for further research. It should also be noted that both pressure and water-vapor density decrease rapidly with increasing altitude, leading to reduced gaseous absorption compared with near-ground horizontal paths. Consequently, microwave, mmWave, and THz links become increasingly favorable for SAGIN nodes operating at higher altitudes.

%%---Fig. 5------
\begin{figure}[t]
\centering
\includegraphics[scale=0.55]{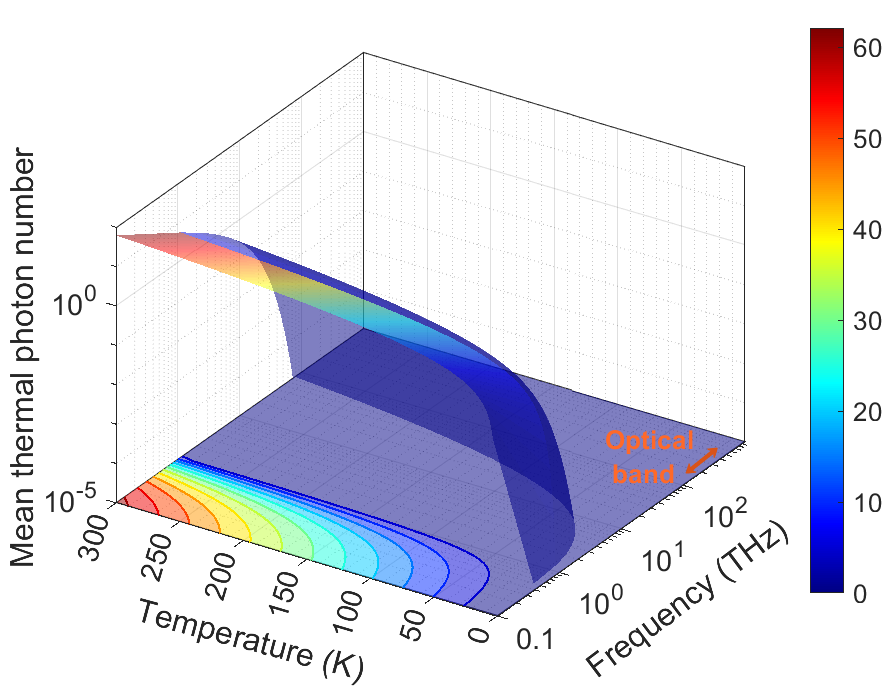}
\caption{Mean thermal photon number as a function of environmental temperature in Kelvin (K) and frequency.}
\label{fig_5}
\end{figure}
%%-------------
As the operating frequency decreases from the optical and infrared bands toward the microwave region, the photon energy becomes too low for efficient generation or detection of individual photons, while the background thermal noise rises sharply \cite{Ottaviani2020,Zhang2023}. Under such conditions, the large population of thermal photons encountered at room temperature overwhelms single-photon signals, making DV-QKD approaches highly impractical because their security depends on the accurate detection of individual photon events with low error probability. This limitation shifts attention toward CV-QKD as a more feasible option for SCQC across the microwave, mmWave, and THz frequency ranges. CV schemes may be regarded as more practical in this regime, because they do not rely on single-photon detection, which is easily masked by thermal background radiation. Instead, they encode information in the field quadratures of coherent or squeezed states, allowing the quantum signal to be extracted through homodyne or heterodyne detection using an LO, effectively averaging over many photon contributions. This collective measurement approach offers a higher SNR and greater resilience to Gaussian thermal noise. Nevertheless, even CV-based systems remain experimentally challenging at room temperature, and secure operation in the microwave band has so far been demonstrated primarily under cryogenic conditions \cite{Fesquet2024}. 

Figure \ref{fig_5} presents the mean thermal photon number as a function of temperature and frequency, based on the formulation in \cite[(3)]{Zhang2023}. It is evident that the optical band typically used for quantum communications, spanning from 193 THz to 375 THz, exhibits an almost negligible thermal photon number, remaining below $10^{-5}$ at room temperatures between 293.15 K and 298.15 K. By contrast, mmWave and lower THz frequencies below 1 THz experience much higher thermal photon occupancy under the same conditions, introducing significant excess noise that can degrade the integrity of quantum signals. This situation creates a fundamental trade-off in system design. Frequencies below 0.5 THz offer reduced atmospheric attenuation, as shown in Fig. \ref{fig_4}, but are heavily affected by thermal noise at room temperatures. On the other hand, frequencies above 1 THz experience lower thermal noise but suffer from strong atmospheric attenuation. A feasible, although costly approach is to operate under cryogenic conditions near absolute 0 K to minimize thermal photon generation \cite{Zhang2023}. However, the associated complexity and cost significantly limit its practical applicability.
%%---------V. Open Problems-----------
\section{Open Problems}
In practical deployments, the choice between SCQC \cite{Pan2025,Qi2016,Kumar2019} and CCQC schemes \cite{Xu2023,Shao2025,Schreier2023} should be made adaptively, based on network infrastructure, performance requirements, and deployment budgets. In this work, we focus on SCQC as a cost- and resource-efficient solution facilitated by reduced hardware components and transceiver sharing. However, its application to multi-band SAGINs still presents significant open problems that require further investigation.
\subsection{\textbf{\underline{Problem 1}:} \textbf{Closing the gap between data rate and key rate in optical SCQC}}
A key limitation in SCQC systems is the mismatch between high classical data rates and relatively low quantum key generation rates. In DV QKD–based SCQC \cite{Pan2025}, classical data is encrypted using a preshared one-time pad key, which must be as long as the message and used only once. This dependence limits sustainability, as the key sink requires continual replenishment to maintain encryption. Although some secret bits can be extracted from quantum measurement results to partially replenish the key sink, this mechanism is only effective under favorable channel conditions. Without effective generation of fresh secret key bits during operation, the secure key sink may be exhausted, compromising continued encryption. In CV QKD–based SCQC \cite{Qi2016}, the data rates are constrained by the speed of shot-noise-limited homodyne detectors required for quantum signal detection. Although homodyne detectors with GHz bandwidths have been demonstrated for CV QKD at optical wavelengths, classical optical coherent communication rates are now approaching Terabits per second. Continued progress in high-repetition-rate light sources, efficient detectors, and optimized error-correction codes is essential to narrow this gap and enhance both data and key rates in SCQC systems.
\subsection{\textbf{\underline{Problem 2}:} \textbf{Mitigating thermal noise and integrating hardware in microwave and mmWave SCQC}}
In the microwave and mmWave bands, the implementation of SCQC is fundamentally challenged by high thermal photon occupancy at room temperature, which introduces substantial excess noise into the quantum channel. As shown in Fig. \ref{fig_5}, frequencies below 300 GHz exhibit elevated mean thermal photon numbers, severely limiting the SNR required for accurate quantum state discrimination. To mitigate this, cryogenic cooling is often employed to mitigate thermal noise, but it significantly increases system complexity and hinders practical deployment in mobile or airborne platforms within SAGINs. Moreover, current mmWave systems remain constrained by the limited availability of the high-performance modulators and coherent detectors needed for quantum-level displacement and shot-noise-limited detection in CV-QKD. Existing modulators are typically optimized for classical signals and do not offer the resolution required for quantum protocols. Likewise, coherent receivers at microwave/mmWave frequencies still face challenges in achieving the stability and sensitivity needed to discriminate weak quantum signals from noise \cite{Fesquet2024}. Integrated transmitter and receiver designs capable of supporting SCQC remain under-developed. Addressing these challenges calls for the co-design of dual-purpose components and robust protocols optimized for the thermal noise environment and hardware limitations of these bands.
\subsection{\textbf{\underline{Problem 3}:} \textbf{Addressing the quantum technology gap in THz SCQC}}
Despite rapid progress in classical THz communications, SCQC implementation in the THz band remains largely unexplored due to the absence of practical quantum-compatible components. Unlike the microwave domain, which has advanced superconducting microwave technologies, or the optical domain having field-tested QKD systems, the THz range remains limited by the lack of mature coherent quantum sources and sensitive detectors capable of operating at low noise levels. This gap in quantum technologies prevents the realization of SCQC protocols that rely on precise state preparation and detection. Although the THz spectrum offers a unique balance between ultra-wide bandwidth and compact hardware integration, progress depends on breakthroughs in materials such as silicon photonics, as well as on the development of compact, high-speed, quantum-capable electronics \cite{Ottaviani2020}. Without these enabling technologies, realizing SCQC in the THz band remains a pressing open research problem.
%%---------VI. Conclusion-----------
\section{Conclusions}
SCQC presents a compelling solution for embedding quantum-secured capabilities into next-generation wireless networks. By enabling the simultaneous transmission of classical and quantum signals over a shared system, it significantly reduces hardware components, lowers deployment cost, and improves energy efficiency. This is particularly important for resource-constrained non-terrestrial platforms within SAGINs. While optical frequencies remain the most mature domain for SCQC, extending these concepts to the microwave, mmWave, and THz bands using CV QKD is essential for realizing SCQC in multi-band SAGINs. However, this expansion faces critical challenges, including high thermal photon noise, strong frequency-dependent atmospheric attenuation, and unresolved issues in hardware development, propagation modeling, and protocol adaptation. Continued research is essential to overcome these limitations and to advance secure unified quantum-classical communications across the full spectrum.
%%---------VI. Acknowledgments-----------
\section*{Acknowledgments}
The work of Shinya Sugiura was supported in part by the Japan Science and Technology Agency (JST) ASPIRE (Grant JPMJAP2345), in part by the JST FOREST (Grant JPMJFR2127), and in part by the Japan Society for the Promotion of Science (JSPS) KAKENHI (Grant 24K21615), The work of Phuc V. Trinh was supported in part by the JSPS KAKENHI (Grant 24K17272) and in part by the Telecommunications Advancement Foundation (TAF). The work of Lajos Hanzo was supported in part by the following Engineering and Physical Sciences Research Council (EPSRC) projects: Platform for Driving Ultimate Connectivity (TITAN) under Grant EP/Y037243/1 and EP/X04047X/1; Robust and Reliable Quantum Computing (RoaRQ, EP/W032635/1); India-UK Intelligent Spectrum Innovation ICON UKRI-1859; PerCom (EP/X012301/1); EP/X01228X/1; EP/Y037243/1.

\section*{Biographies}
\vspace{-1cm}
\begin{IEEEbiographynophoto}{Phuc V. Trinh} (Senior Member, IEEE) (trinh@iis.u-tokyo.ac.jp) received the Ph.D. degree in computer science and engineering from The University of Aizu, Japan, in 2017. Since 2023, he has been a Project Research Associate with the Institute of Industrial Science, The University of Tokyo, Japan. His current research interests include optical and wireless communications for space, airborne, and terrestrial networks.
\end{IEEEbiographynophoto}

\begin{IEEEbiographynophoto}{Shinya Sugiura} (Senior Member, IEEE)  (sugiura@iis.u-tokyo.ac.jp) received the Ph.D. degree from the University of Southampton, U.K., in 2010. Since 2018, he has been with the Institute of Industrial Science, The University of Tokyo, Japan, where he is currently a full Professor. He is an Editor for IEEE TWC, IEEE TCOM, and IEEE WCL.
\end{IEEEbiographynophoto}

\begin{IEEEbiographynophoto}{Carlo Ottaviani} (carlo.ottaviani@york.ac.uk) is a Senior Lecturer at the Department of Computer Science and York Centre for Quantum Technologies at the University of York, U.K. Over the years his research interests have ranged from non-linear atom-optics to quantum information \& communication to post-quantum cryptography. He contributed to pioneering practical applications on the use of continuous variables and quantum channels for cryptography and on the use of photons as carriers of quantum information. His research focuses on quantum technologies and in particular quantum cryptography, quantum networks and post-quantum cryptography.
\end{IEEEbiographynophoto}

\begin{IEEEbiographynophoto}{Chao Xu} (Senior Member, IEEE) (cx1g08@ecs.soton.ac.uk) received the Ph.D. degree from the University of Southampton, U.K., in 2015. He is currently a Senior Lecturer with Next Generation Wireless Research Group, University of Southampton. He was the recipient of 2023 Marie Sklodowska-Curie Actions Global Postdoctoral Fellowships with the highest evaluation score of 100/100.
\end{IEEEbiographynophoto}

\begin{IEEEbiographynophoto}{Lajos Hanzo} (Life Fellow, IEEE) (lh@ecs.soton.ac.uk) received Honorary Doctorates from the Technical University of Budapest (2009) and Edinburgh University (2015). He is a Foreign Member of the Hungarian Science-Academy, Fellow of the Royal Academy of Engineering (FREng), of the IET, of EURASIP and holds the IEEE Eric Sumner Technical Field Award. For further details please see \url{http://www-mobile.ecs.soton.ac.uk}, \url{https://en.wikipedia.org/wiki/Lajos_Hanzo}.
\end{IEEEbiographynophoto}
\balance
\newpage

\vfill

\end{document}